\documentclass{article}

\usepackage{arxiv}

\usepackage[utf8]{inputenc} % allow utf-8 input
\usepackage[T1]{fontenc}    % use 8-bit T1 fonts
\usepackage{hyperref}       % hyperlinks
\usepackage{url}            % simple URL typesetting
\usepackage{booktabs}       % professional-quality tables
\usepackage{amsfonts}       % blackboard math symbols
\usepackage{nicefrac}       % compact symbols for 1/2, etc.
\usepackage{microtype}      % microtypography
\usepackage{graphicx}
\usepackage{natbib}
\usepackage{doi}
\usepackage{bm}
\usepackage{amsmath}

\title{Nonparametric tests for interaction in two-way ANOVA with balanced replications}

\date{October 6, 2024}

\author{ \href{https://orcid.org/0009-0009-2895-204X}{\includegraphics[scale=0.06]{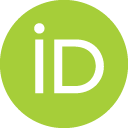}\hspace{1mm}Bao Khue Tran}\\ 
	Department of Mathematics and Statistics\\
	Kenyon College\\
	Gambier, OH \\
	\texttt{tran3@kenyon.edu} \\
	\And
        Amy S.~Wagaman \\
	Department of Mathematics and Statistics\\
	Amherst College\\
        Amherst, MA\\
	 \And
	Andrew Nguyen \\
	  Department of Statistics and Operations Research\\
        University of North Carolina at Chapel Hill\\
        Chapel Hill, NC\\
	 \And
	 David Jacobson \\
      Department of Mathematics and Statistics\\
      Amherst College\\
      Amherst, MA\\
	\And
	Bradley Hartlaub \\
	  Department of Mathematics and Statistics\\
	Kenyon College\\
	Gambier, OH \\
}

\begin{document}
\maketitle

\begin{abstract}
Nonparametric procedures are more powerful for detecting interaction in two-way ANOVA when the data are non-normal. In this paper, we compute null critical values for the aligned rank-based tests ($APCSSA/APCSSM$) where the levels of the factors are between 2 and 6. We compare the performance of these new procedures with the ANOVA \textit{F}-test for interaction, the adjusted rank transform test ($ART$), Conover’s rank transform procedure ($RT$), and a rank-based ANOVA test ($\texttt{raov}$) using Monte Carlo simulations. The new procedures $APCSSA/APCSSM$ are comparable with existing competitors in all settings. Even though there is no single dominant test in detecting interaction effects for non-normal data, nonparametric procedure $APCSSM$ is the most highly recommended procedures for Cauchy errors settings.
\end{abstract}


\section{Introduction}
Factorial designs allow researchers to explore main effects and interactions. Particularly, detecting interaction is crucial to conducting data analysis as the presence of interaction in the data can influence whether a factor is significant or not. In a parametric setting, the usual test for interaction is an $F$-test \citep{Montgomery2020}. However, for data collected in various fields and industries, errors tend to not follow the normal distribution. The use of tests with known issues provides continued motivation for developing appropriate nonparametric tests for interaction in the two-way layout with multiple replications per cell. We consider the model for the general two-way layout and some different types of interaction. We focus on a balanced design with an equal number of replications per cell.

In the case of a two-way layout with balanced replications per cell, the general model is
\begin{equation}
Y_{ijk} = \alpha_i + \beta_j + \gamma_{ij} + \varepsilon_{ijk}, \;
i = 1,...,I, \; j = 1,...,J, \mbox{and} \; k = 1,...,K,
\label{model}
\end{equation}
where two factors $U$ and $V$, having $I$ and $J$ levels respectively, are being investigated and $K$ is the number of replications per cell. The $Y_{ijk}$'s are $IJK$ observations which are mutually independent random variables. The error terms, denoted by the $\varepsilon_{ijk}$'s, are assumed to have a common median $\theta$. For notation, $\alpha_i$ is the effect of the $i^\text{th}$ level of factor $U$, $\beta_j$ is the effect of the $j^\text{th}$ level of factor $V$, and $\gamma_{ij}$ is the effect of the interaction between the $i^\text{th}$ level of factor $U$ and the $j^\text{th}$ level of factor $V$. \\
The hypotheses for the typical test of interaction, with no restrictions on $\alpha_i$ and $\beta_j$, are
\begin{align*}
    H_0&: \gamma_{ijk} = 0, i=1, \dots, I, j=1, \dots, J, \mbox{ and } k=1, \dots, K;\\
    H_A&: \; \gamma_{ijk} \mbox{'s not all zero.}
\end{align*}

Stating that not all $\gamma_{ijk}$ values are zero is only one way of stating that interaction is present. Multiple definitions of interaction have been developed, and some are only applicable to the location-family model, while others are more general. 

In this paper, we propose new aligned rank-based tests ($APCSSA/APCSSM$) for interaction in the general two-way layout with balanced replications per cell. We start with a review of existing test procedures for interaction. Then, we set forth our procedures. Through Monte Carlo simulation power studies with seven competing tests for interaction, we demonstrate the consistent performance of $APCSSA/APCSSM$ in various settings. We conclude with a discussion of our findings and thoughts on future work in this area.

\section{Review of test procedures for interaction}
\label{sec:review}

Researchers have developed a variety of procedures to test for interaction in the general two-way layout. The classical test procedure for determining the presence of interaction (as defined by non-zero $\gamma_{ijk}$'s) in a parametric setting is the standard ANOVA $F$-test \citep{Montgomery2020}. When the error terms in (\ref{model}) are normally distributed, the two-way analysis of variance sum of squares identity partitions the variability and leads to an $F$-test for our null hypothesis of no interaction. The $F$-test assumptions of normality and common variance may not be met in all situations, however. When the assumptions necessary for performing an $F$-test are not met, a nonparametric test should be used.

Multiple nonparametric tests for interaction exist in the literature. Conover and Iman (\citeyear{Conover1976}) suggest using the rank transform approach ($RT$) in factorial settings. $RT$ uses joint ranks (averages for ties correction), and then uses the parametric $F$-test on the ranks. Another nonparametric test for interaction based on the rank transform approach is the aligned rank transform ($ART$). With the $ART$ procedure, an alignment is performed in the rows and columns before the joint ranking of the response variables. Wobbrock, Findlater, Gergle, and Higgins (\citeyear{Wobbrock2011}) align the data before performing analysis like $RT$ to study individual effects. Alignment is performed by computing the row and column averages and subtracting each from all corresponding entries in the individual rows and columns. After obtaining the joint ranks on the aligned data, the $F$-statistic is calculated on the joint ranks. 

The aligned rank transform has been found to perform better than the rank transform, but it still has problems with elevated Type I error rates and with nonnormal error terms \citep{Richter1999, Luepsen2017}. Mansouri and Chang (\citeyear{Mansouri1995}) performed a comparative analysis between the $F$-test, rank transform, and two versions of the aligned rank transform. They conclude that the aligned rank tests are preferred over the other test procedures. They also note that all the test procedures they considered performed poorly in the Monte Carlo power study with errors drawn from a Cauchy distribution. In our power studies, we use the $\texttt{art}$ procedure from the $ARTool$ package in $\mathsf{R}$ \citep{Kay2021}.

Salazar-Alvarez, Tercero-Gómez, Cordero-Franco, and Conover (\citeyear{Salazaralvarez2014}) conduct a literature review and conclude that most of the techniques were based on $RT$. In order to provide a more complete comparison of nonparametric tests for interaction, we will consider methods that do not stem from Conover and Iman's $RT$.

A different testing approach was proposed by De Kroon and Van Der Laan (\citeyear{Dekroon1981}), based on their definition of rank interaction. Their test statistic was developed to detect the presence of rank interaction in the two-way layout with multiple replications per cell. As noted by De Kroon and Van Der Laan (\citeyear{Dekroon1981}) and Hartlaub et al. (\citeyear{Hartlaub1999}), the proposed statistic only works well in detecting a rank interaction of type $U^{*}(V)$ if no main effect for $U$ is present (and vice versa for detecting type $V^{*}(U)$, which works well where no main effect for $V$ is present). 

Another way to detect interaction in factorial design is through the lens of linear regression, specifically rank-based linear models. Kloke and McKean (\citeyear{Kloke2012}) proposed a rank-based analysis for all three hypotheses including the main effects and interactions based on a reduction of dispersion from the reduced to the full model. For the computations in this paper, we have utilized the $\texttt{raov}$ function in $\texttt{Rfit}$ package to compute their test statistic for interaction \citep{Kloke2012}.

Lastly, Hartlaub, Dean, and Wolfe (\citeyear{Hartlaub1999}) developed procedures to test for interaction in the two-way layout with one observation per cell. An invariance problem with their statistics was solved by Lehman, Wolfe, Dean, and Hartlaub \citep{Lehmann2001} who proposed symmetrized procedures. The symmetrized procedures, $S-SA$ (symmetrized statistics aligned by averages) and $S-SM$ (symmetrized statistics aligned by medians) are based on the statistics $CRA$ and $RCA$, and $CRM$ and $RCM$, respectively. In short, these procedures align within the rows or columns using averages or medians, and then rank within the other dimension. Reversing the aligning and ranking to create two statistics is similar to checking for both row and column concordance. The ranks are then combined in a cross-comparison framework to form appropriate statistics to test for interaction. A challenge for this procedure is that null means and variances for the statistics must be computed before the significance of the test statistic may be determined.

\section{Nonparametric test for interaction proposal}
\label{sec:proposal}

Salazar-Alvarez, Tercero-Gómez, Cordero-Franco, and Conover (\citeyear{Salazaralvarez2014}) recommend developing new nonparametric methods that are not based on $RT$ to detect interaction. The test statistics proposed by Hartlaub, Dean, and Wolfe (\citeyear{Hartlaub1999}) were found to perform well in the two-way layout with a single replication per cell.  Thus, we propose an extension of their technique for the case of balanced replications per cell. Our proposed statistics use the technique of crossed comparisons \citep{Tukey1991} to detect the interactions. Initial investigations summarizing cell information into a single statistic (such as a median or mean) and then applying the methods from Hartlaub, Dean, and Wolfe (\citeyear{Hartlaub1999}) did not perform as well as these extended methods where all possible comparisons were examined. 

Our proposed statistics generalize the comparison idea with all possible comparisons. We eliminate nuisance effects by aligning with averages or medians to remove one of the nuisance effects (row or column) and ranking within the columns or rows to remove the other. We call the proposed statistics $APCSSA$ and $APCSSM$. The names of the statistics come from the idea that they are extensions of $S\text{-}SA$ and $S\text{-}SM$ from Hartlaub, Dean, and Wolfe \citeyear{Hartlaub1999}, where we have added APC to stand for all possible comparisons. 

Next, we describe how our proposed statistics, $APCSSA$ and $APCSSM$, are computed. We begin with $APCSSA$, the all possible comparisons extension of $SSA$. $APCSSA$ is the maximum of two standardized statistics, so we begin by outlining their calculation.  

Step 1. Calculate $APCCRA$, which stands for all possible comparisons (APC), column alignment (C), row ranking (R), using averages for the alignment (A). Again, the name of the statistic just reflects that we are aligning within the columns using column averages, and then ranking within the rows, and then using all possible comparisons to create the statistic (below). We compute the $J(J-1)/2$ crossed comparisons denoted $V_{jj\;'}$.
\begin{equation}
V_{jj\;'} = \sum_{1\leq \:i<}\sum_{i\;'\leq \: I}\; \; \sum^K_{k_1
= 1}\sum^K_{k_2 = 1}\sum^K_{k_3 = 1}\sum^K_{k_4 = 1} \left\{
\left( r_{ijk_1} + r_{i\;'j\;'k_2} \right) - \left(r_{i\;'jk_3} +
r_{ij\;'k_4} \right) \right\}^2.
\label{Crossedcompcalc}
\end{equation}

\noindent{$APCCRA$ is the maximum of the $V_{jj\;'}$'s.}

Step 2. Calculate $APCCRAD$, which is just a scaled version of $APCCRA$. Divide $APCCRA$ by $K^4I(I-1)/2$, the number of summands in $APCCRA$, to obtain this scaled version of the crossed comparisons for the maximum column comparison. That is,
\begin{equation}
APCCRAD = \frac{APCCRA}{K^4I(I-1)/2}.
\label{APCCRADcalc}
\end{equation}

Step 3. Calculate $APCRCA$, which stands for all possible comparisons (APC), row alignment (R), column ranking (C), using averages for the alignment (A). Repeat Step 1, with alignment in the rows and ranking in the columns. $APCRCA$ is computed by taking the maximum of $I(I-1)/2$ possible row comparisons.

Step 4. Calculate $APCRCAD$.  $APCRCAD$ is computed by dividing $APCRCA$ by $K^4J(J-1)/2$.

Step 5. Standardization. $APCCRAD$ and $APCRCAD$ are further standardized by subtracting the appropriate null mean and dividing by the appropriate null standard deviation. In order to do this, one must find
\begin{equation}
APCCRAD^* = \frac{APCCRAD - E_0(APCCRAD)}{\sqrt{V_0(APCCRAD)}}, \\
\label{standardizing1}
\end{equation}
\begin{equation}
\mbox{and} \; APCRCAD^* = \frac{APCRCAD -
E_0(APCRCAD)}{\sqrt{V_0(APCRCAD)}},
\label{standardizing2}
\end{equation}
where $E_0(APCCRAD)$ and $E_0(APCRCAD)$ are the null means of $APCCRAD$ and $APCRCAD$ respectively, and $V_0(APCCRAD)$ and $V_0(APCRCAD)$ are the null variances of $APCCRAD$ and $APCRCAD$ respectively. These null means and variances are computed via simulation and their values are available on Github at \textit{\href{https://github.com/tranbaokhue/Rank-based-InteractionTest}{https://github.com/tranbaokhue/Rank-based-InteractionTest}}. Note that if $I=J$, then by symmetry, the null means and null variances are equal and only one set needs to be computed. 

Step 6. Calculate $APCSSA$. $APCSSA$ is the maximum of $APCCRAD^*$ and $APCRCAD^*$.\\
\noindent
Alternatively, using medians to align the data instead of averages in our procedure yields $APCSSM$. The common ties correction of using average ranks for ties should be used during the ranking process.

\section{Simulations}
\label{sec:sims}
\subsection{Simulation settings and notes}

With the increase in $\mathsf{R}$ packages used to analyze two-way ANOVA models, Feys (\citeyear{Feys2016}) cautions researchers against choosing tests for their $p$-values based on a given dataset. In order to compare the proposed statistics with existing competitors, we performed a simulation power study with the $F$, $RT$, $ART$, $DEKR$, $\texttt{raov}$ statistics, and our two proposed statistics $APCSSA$ and $APCSSM$. Multiple settings, with factor levels from 2 to 6 and replications between 2 and 9, were investigated and in this section we show selected results from four settings: $3 \times 2 \times 3$, $3 \times 3 \times 3$, $3 \times 4 \times 2$, and $4 \times 6 \times 2$, where the format $I \times J \times K$ gives the number of levels for each factor and the number of replications per cell.

The null distributions for $APCSSA/APCSSM$ were derived for all settings with 2 to 6 levels in each main factor and the number of replications per cell ranging from 1 to 5 using Monte Carlo simulation. The null distributions were based on 100,000 computations of the statistics with no interaction or main effects present and using normal error terms. In cases where symmetrization was needed, e.g. $APCSSA$ and $APCSSM$, two sets of 100,000 computations were used. The first set of 100,000 was used to determine expected values and variances for use in the symmetrizations, and the second set for the null distribution determination of the symmetrized statistics. During the ranking process, the common ties correction of using average ranks was employed for the our new proposed procedures, $APCSSA$ and $APCSSM$. Expected values and variances as well as critical values for our proposed statistics are available through Github (\textit{\href{https://github.com/tranbaokhue/Rank-based-InteractionTest}{https://github.com/tranbaokhue/Rank-based-InteractionTest}}). 

We used Monte Carlo simulation to perform the power comparisons for all statistics at the $0.05$ significance level. Multiple main effects were chosen for each factor, and two different types of interaction were studied. These types are product and specific interaction. Product interaction is interaction where the $\gamma_{ijk}$'s are related to the main effects, simply as $\gamma_{ijk}= \lambda \alpha_i\beta_j$, where $\lambda$ is a general scaling factor (Tukey, 1949). In specific interaction, $2 \times 2$ matrices
$$\begin{bmatrix}
c & -c\\
-c & c
\end{bmatrix}$$
where $c \in \mathbb{R}$, are embedded in the rows or columns. Details on the main effects and interaction effects examined are available through Github at \textit{\href{https://github.com/tranbaokhue/Rank-based-InteractionTest}{https://github.com/tranbaokhue/Rank-based-InteractionTest}}. We note that main effects are absent for each factor at factor level 1. 

When generating the data, error terms were drawn from the $\text{Normal}(0, 1)$, $\text{Uniform}(-2, 2)$, $\text{Exponential}(1)$, $\text{Double Exponential}\left(0, \frac{1}{\sqrt{2}}\right)$, or $\text{Cauchy}(0, 1)$ distributions. For each combination of main effects, interaction effects, and error terms, we conduct 10,000 simulations for power comparisons. We use varying interaction effects and no interaction to attain the full power curves for each procedure under each setting. All simulations were computed with R Statistical Software (v.4.3.1; R Core Team 2023).

\subsection{Simulation results}

Our selected power comparison results are shown via figures. Figures 1 and 4 are from the $3 \times 4 \times 2$ setting, while Figures 3 and 5 are from the $3 \times 2 \times 3$ setting and $3 \times 3 \times 3$ setting. Finally, Figures 2 and 6 are from the  $4 \times 6 \times 2$ setting. While Figures 1, 3, 4, and 5 include power curves from settings with product interaction, Figures 2 and 6 present the powers of the tests when the $2 \times 2$ matrices of specific interaction are embedded in the first two columns of the simulated data. Both interaction types as well as all five distributions of errors are shown in our selected results.

In Figure 1, with normal errors, even thought $ART$ has slightly higher power than the $F$-test, $ART$ suffers from inflated Type I error. Thus, we still recommend the $F$-test for normal error data compared to the others (in order from second-best to worst, $APCSSA$, $\texttt{raov}$, $APCSSM$, $RT$, and $DEKR$). Our new proposed procedure $APCSSA$ is comparable to the $F$-test in terms of power. We also quickly see that the $DEKR$ statistic does not do well when both factors have main effects. 

In Figure 2, looking at uniform errors, we can see that the main effects clearly nullified $DEKR$ interaction detection. We can see that $APCSSA$ performs the best followed closely by the $F$-test and others.

\begin{figure}
\includegraphics[width=\linewidth]{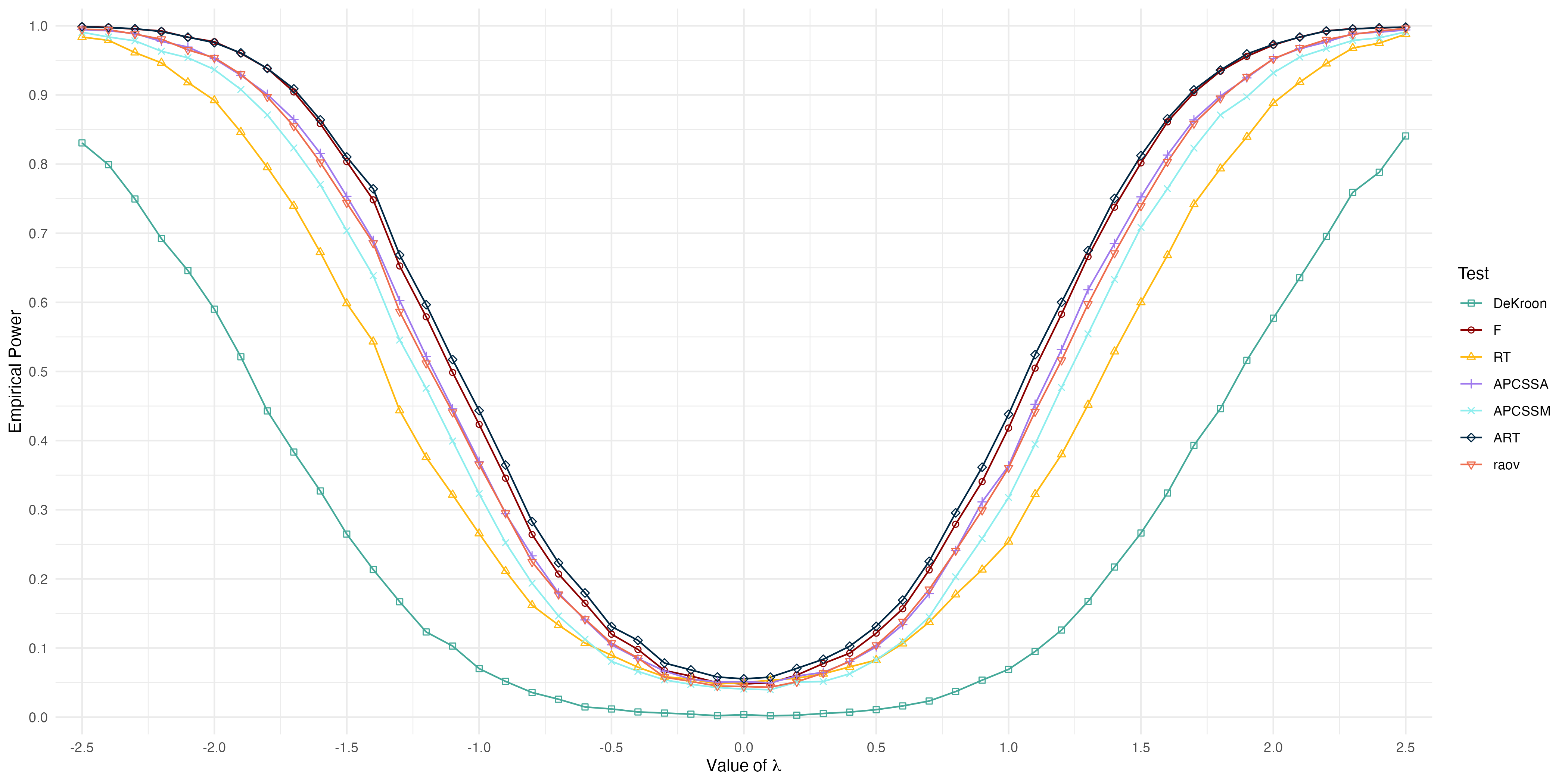}
\caption{Power curves in the $3 \times 4 \times 2$ setting with product interaction ($\gamma_{ijk}=\lambda \alpha_i \beta_j$), $\bm{\alpha}=(-1, 0, 1)$, $\bm{\beta}=(-1, -0.5, 0.5, 1)$, and standard normal errors.}
\end{figure}

\begin{figure}
\includegraphics[width=\linewidth]{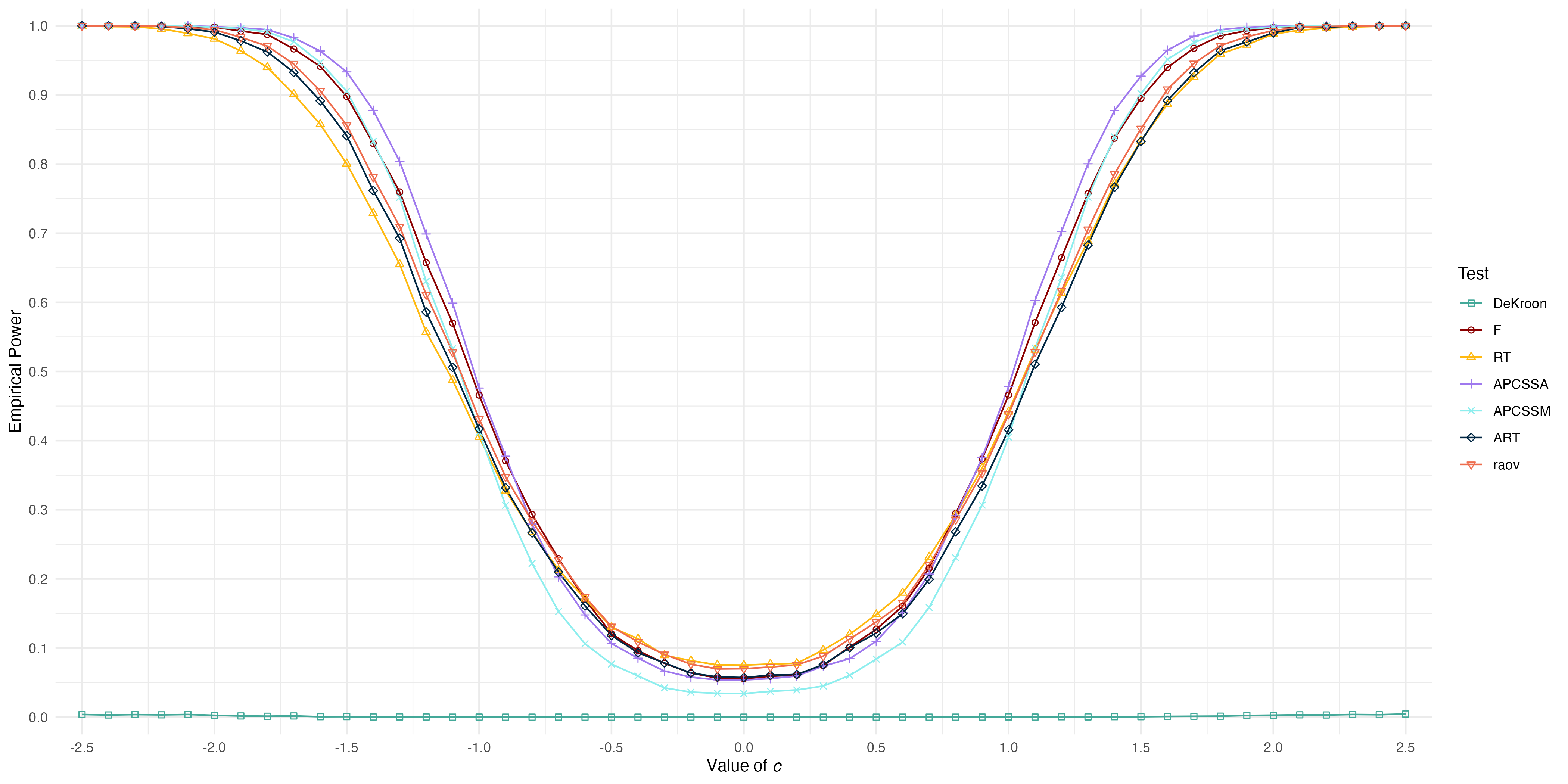} %SI in two columns
\caption{Power curves in the $4 \times 6 \times 2$ setting with specific interaction, $\bm{\alpha}=(-2, 0, 0.5, 1.5)$, $\bm{\beta}=(-1.5, -1, -0.5, 0.5, 1, 1.5)$, and uniform errors.}
\end{figure}

\pagebreak 

In Figure 3, when the data follow an exponential distribution, $\texttt{raov}$ is the most powerful but it has an extremely elevated Type I error. Following closely behind are the $ART$ and the proposed $APCSSA$. These two are powerful, with only slightly inflated significance levels. Even though $APCSSA$ is comparable, we would generally recommend comparing with the $F$-test results. Facing double exponential errors, in Figure 4, we can see that $ART$ is the most powerful, but $APCSSA$, $\texttt{raov}$, and the $F$-test all have roughly the same empirical powers. From both these figures, it is clear how $RT$ and $DEKR$ are easily influenced by main effects.

\begin{figure}
\includegraphics[width=\linewidth]{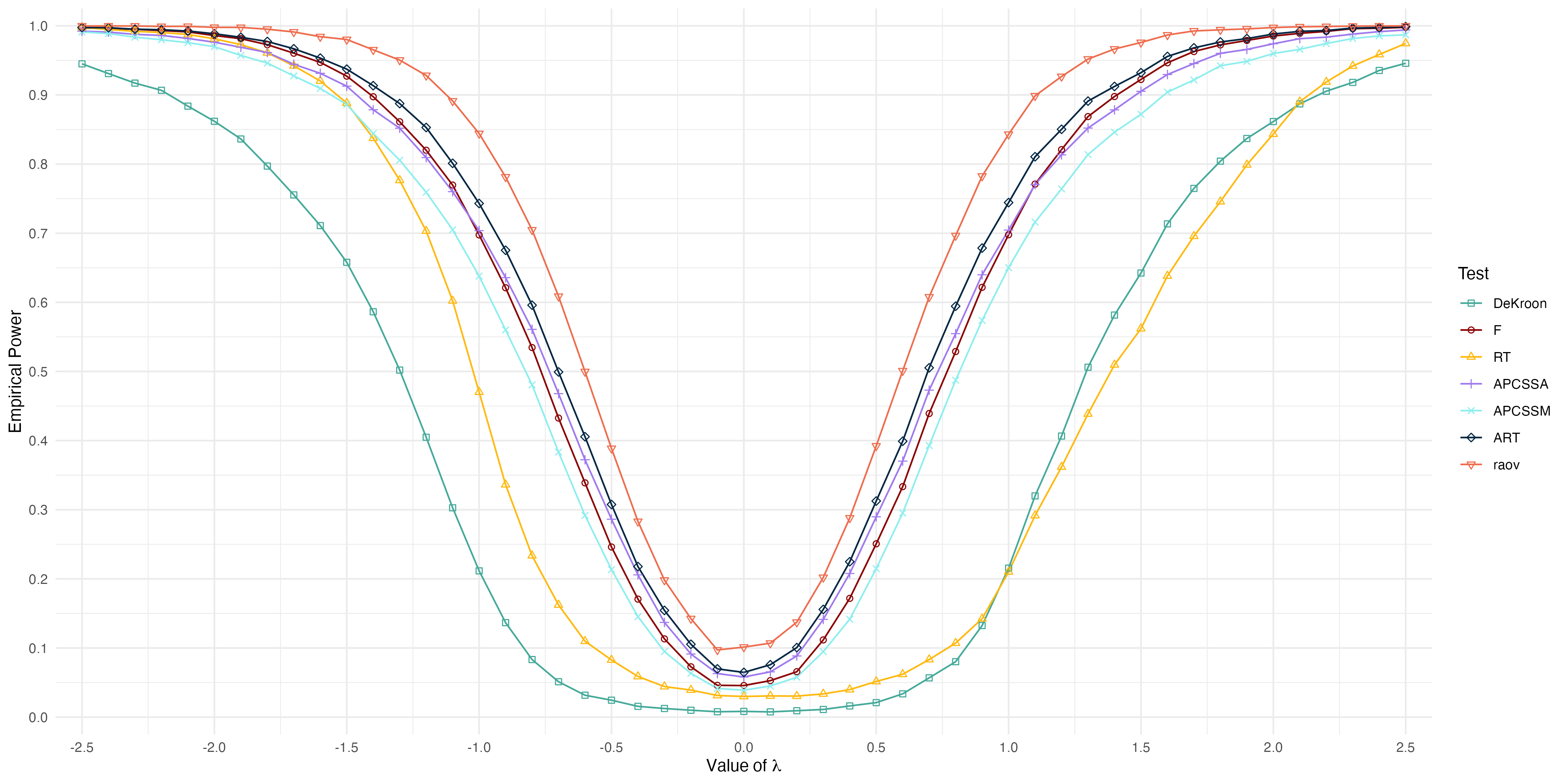}
\caption{Power curves in the $3 \times 2 \times 3$ setting with product interaction ($\gamma_{ijk}=\lambda \alpha_i \beta_j$), $\bm{\alpha}=(-0.5, -0.5, 1)$, $\bm{\beta}=(-1,1)$, and exponential errors.}
\end{figure}

\begin{figure}
\includegraphics[width=\linewidth]{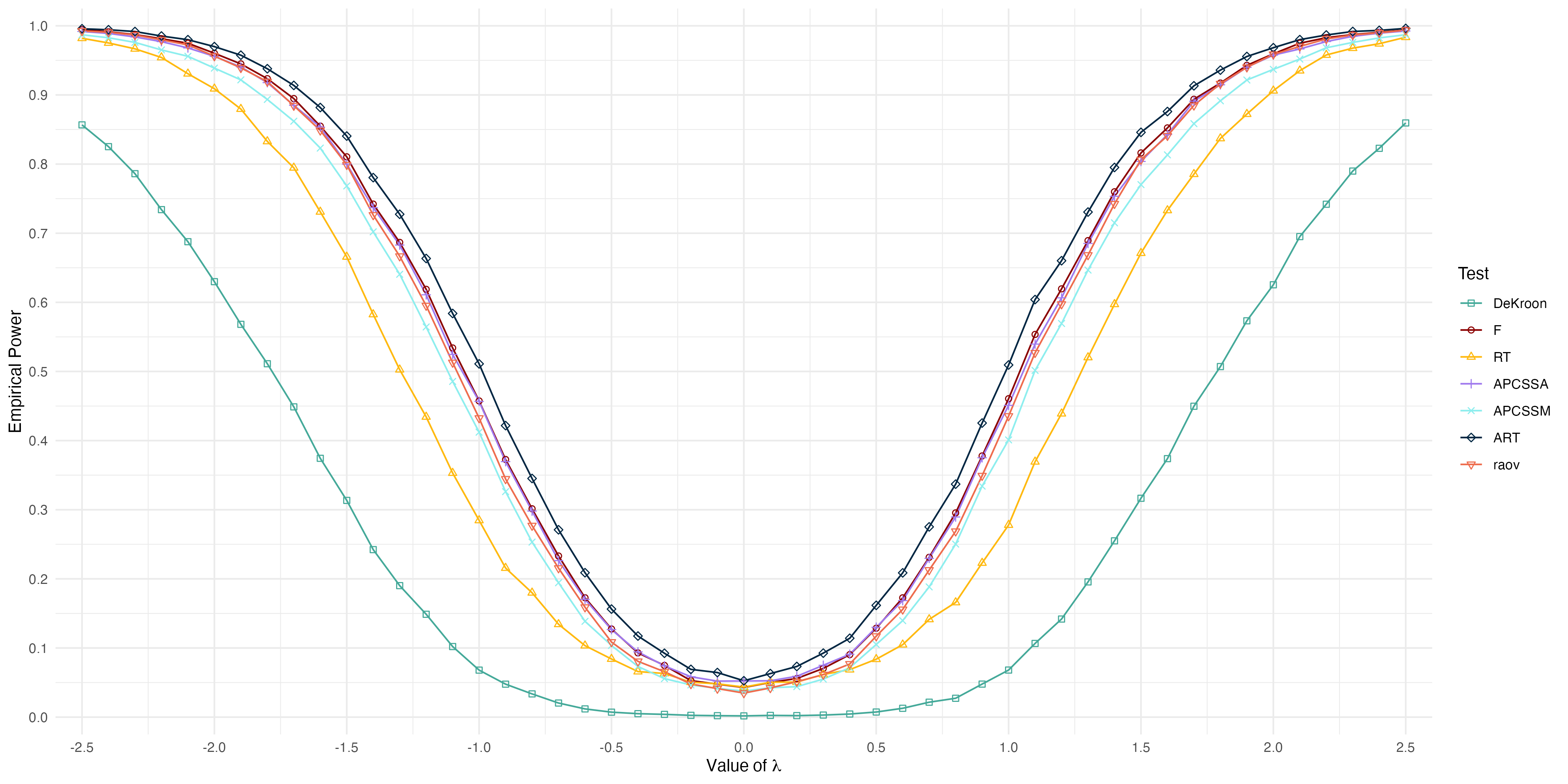}
\caption{Power curves in the $3 \times 4 \times 2$ setting with product interaction ($\gamma_{ijk}=\lambda \alpha_i \beta_j$), $\bm{\alpha}=(-1, 0,1)$, $\bm{\beta}=(-1, -0.5, 0.5, 1)$, and double exponential errors.}
\end{figure}

\pagebreak

Finally, as seen in Figure 5 and Figure 6, when faced with the challenges of Cauchy errors, our proposed test $APCSSM$ performs the best in various settings (both product interaction and specific interaction). Kloke and McKean's $\texttt{raov}$ might be more powerful, but when there is no interaction, the rejection rate remains way over 0.1. We can also see how the $F$-test, $DEKR$, $APCSSA$, and $ART$ perform when the data have many outliers.

\begin{figure}
\includegraphics[width=\linewidth]{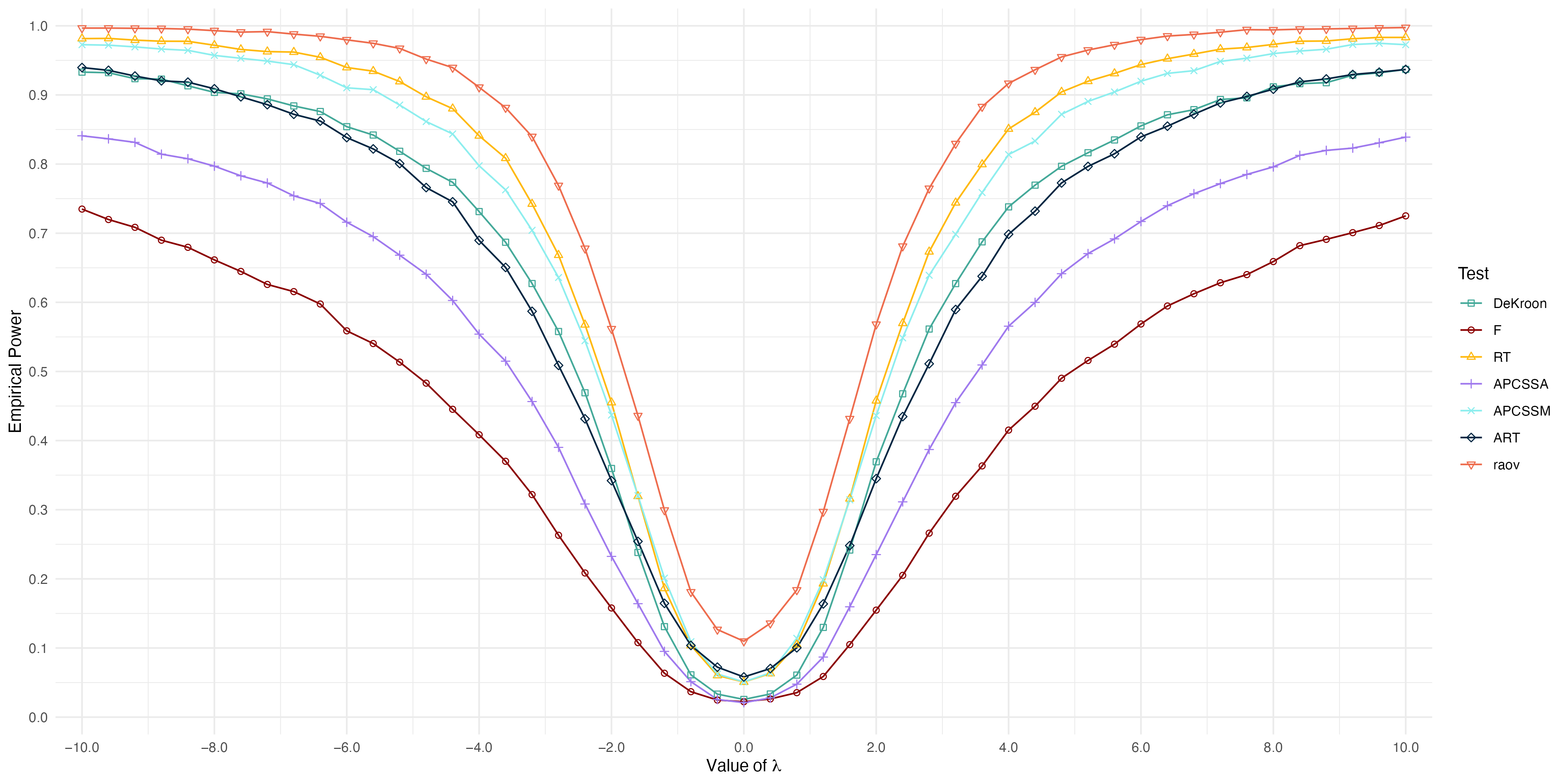}
\caption{Power curves in the $3 \times 3 \times 3$ setting with product interaction ($\gamma_{ijk}=\lambda \alpha_i \beta_j$), $\bm{\alpha}=(-1,0,1)$, $\bm{\beta}=(-1,0,1)$, and Cauchy errors.}
\end{figure}

\begin{figure}
\includegraphics[width=\linewidth]{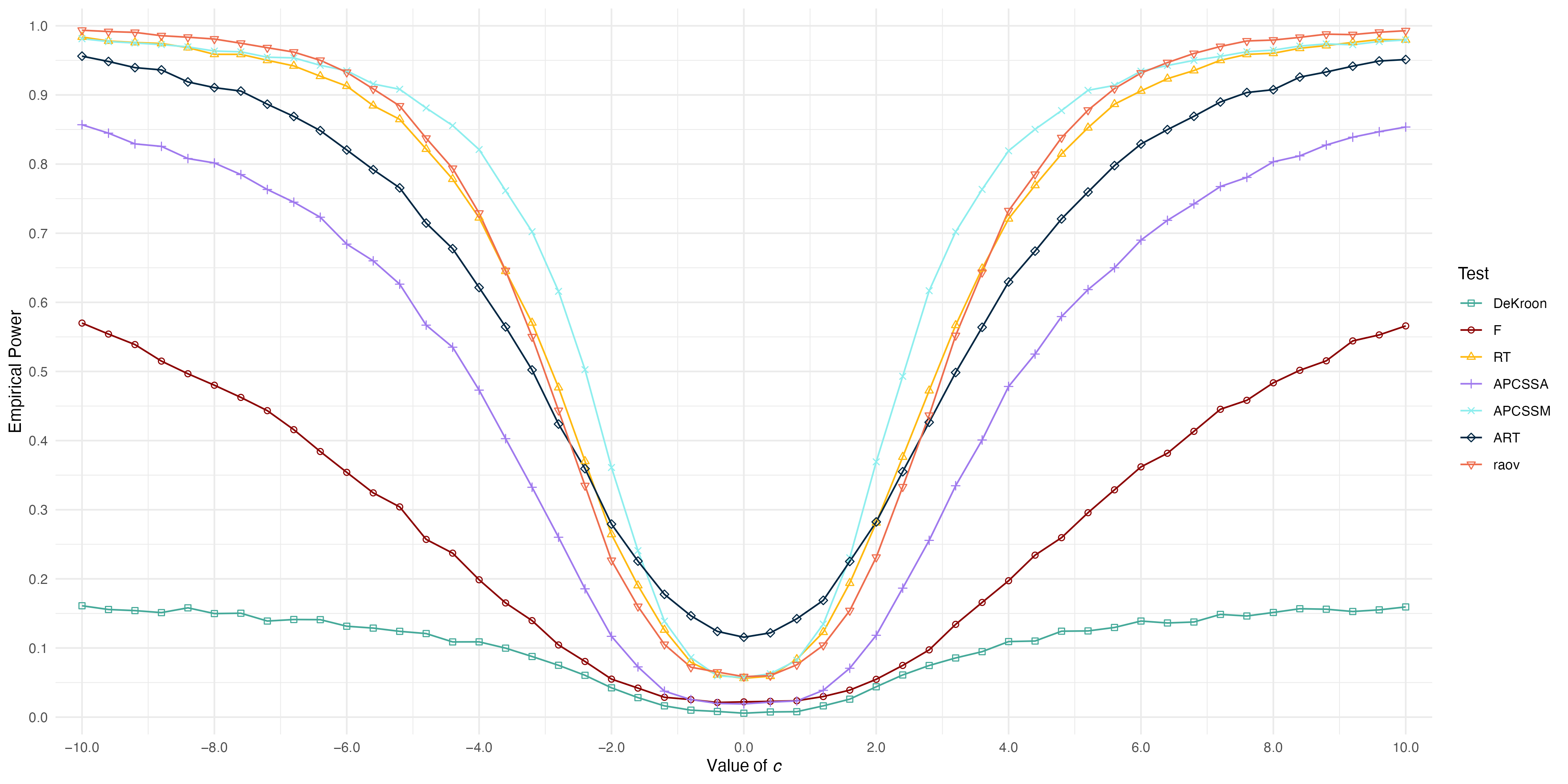} %SI in two columns
\caption{Power curves in the $4 \times 6 \times 2$ setting with specific interaction, $\bm{\alpha}=(-2, 0, 0.5, 1.5)$, $\bm{\beta}=(-1.5, -1, -0.5, 0.5, 1, 1.5)$, and Cauchy errors.}
\end{figure}

Our results for other combinations of settings, main effects, interactions and error terms are consistent with these selected results. In summary, the new proposed statistics perform outstanding in situations with Cauchy or double exponential errors, so we advocate their use to detect interaction in these settings.

\pagebreak

\section{Discussion and Future Work}
\label{futurework}

Our simulation studies have verified previous work that $DEKR$ suffers from the introduction of row main effects and is not recommended for interaction detection in the balanced replications per cell setting. With Conover's $RT$ approach, the resulting statistic does not compete well with the $F$-test and suffers elevated Type I error rates when the error terms come from nonnormal distributions. Although Kloke and McKean's $\texttt{raov}$ and $ART$ are much more powerful than $DEKR$ and $RT$, their rejection rates in various settings can be twice the significance level of 0.05. Our proposed statistics $APCSSA$ and $APCSSM$ perform well in settings with exponential and Cauchy error terms respectively. Additionally, when using $APCSSA$, not much power is lost compared to the $F$-test when error terms are from the normal or uniform distributions and $APCSSM$ is undeniably the best procedure for detecting interactions for data with Cauchy errors. In conclusion, we are able to recommend using $APCSSA$ and $APCSSM$ to detect interaction in the two-way layout with balanced replications per cell.

While we have demonstrated that our statistics work well to detect interaction in these settings, future work in this area remains. Further validation of these results with the power comparisons from settings with more replications per cell may be enlightening. Additionally, we have work in progress to develop $\mathsf{R}$ code so that the new statistics, which are computationally intensive, may be easily accessed by interested researchers. It is also natural to consider extending these techniques to the general two-way layout with an unequal number of replications per cell.

\noindent
\Large
\textbf{Acknowledgments}

\noindent
\normalsize
We are grateful to Jessica Jeong for her meticulous verification of $\mathsf{R}$ code and scripts. We would like to thank Amherst College and Kenyon College for supporting our summer research projects.

\bibliographystyle{abbrvnat}
\bibliography{apcRef}

\end{document}